\documentclass[fleqn,10pt,authorversion,onecolumn]{wlscirep}

\usepackage{amsmath, amssymb, amsfonts}
\usepackage{graphicx}
\usepackage{color}
\usepackage{float}
\usepackage{wrapfig}
\usepackage{array}
\usepackage{hyperref}
\usepackage[english]{babel}
\usepackage[utf8]{inputenc}
\usepackage{amsmath}
\newcommand{\fref}[1]{Figure \ref{#1}}
\renewcommand{\vec}[1]{\mathbf{#1}}
\newcommand{\hatt}[1]{\mathbf{\widehat{#1}}}
\hypersetup{
    colorlinks = false,
   urlcolor={1 1 1}
    }

\graphicspath{{figures/}}

\title{Collective gradient sensing in fish schools}

\author[1,*]{James G. Puckett}
\author[1]{Aawaz R. Pokhrel} 
\author[1]{Julia A. Giannini} 

\affil[1]{Department of Physics, Gettysburg College, Gettysburg, Pennsylvania 17325, USA}

\affil[*]{jpuckett@gettysburg.edu}

\begin{abstract}
Throughout the animal kingdom, animals frequently benefit from living in groups.  Models of collective behaviour show that simple local interactions are sufficient to generate group morphologies found in nature (swarms, flocks and mills).   However, individuals also interact with the complex noisy environment in which they live.  In this work, we experimentally investigate the group performance in navigating a noisy light gradient of 
two unrelated freshwater species:  golden shiners ({\it{Notemigonus crysoleucas})} and rummy nose tetra ({\it{Hemigrammus bleheri}}).  We find that tetras outperform shiners due to their innate individual ability to sense the environmental gradient.  Using numerical simulations, we examine how group performance depends on the relative weight of social and environmental information.  Our results highlight the importance of 
balancing of social and environmental information to promote optimal group morphologies and performance.
\end{abstract}
\begin{document}

\flushbottom
\maketitle
%
%
\thispagestyle{empty}

\section*{Introduction}

Collective animal behaviour arises from self-organising social interactions among individuals\cite{Couzin2003,Sumpter2010}. 
While the functional form of these interactions is not trivial to determine from experiments\cite{Puckett2014}, models have shown that simple local social interactions consisting of rules such as repulsion, alignment, and attraction are sufficient to generate observed group morphologies, such as swarms, flocks, and mills\cite{Reynolds1987,Huth1994,Couzin2002,Gautrais2008,HerbertRead2011,Ward2016}.  
However, in nature, individuals must balance social information with individually acquired environmental information\cite{Ward2008,Couzin2009,Miller2013}.  

Social interactions benefit group members in diverse ways. 
Living in groups has been shown to increase foraging ability \cite{Pitcher1982,Bazazi2012} and reduce predation risk
\cite{Partridge1982} through collective vigilance \cite{Taraborelli2012} or escape waves\cite{Herbert-Read2016}.  
Individuals can also use social information to help navigate noisy environmental gradients~\cite{Larkin1969,Grunbaum1998,Berdahl2013}.  
In this case where individuals may benefit from pooling information to overcome inaccurate estimates, which is often called `wisdom of the crowd'\cite{Krause2010} or the `many wrongs' principle \cite{Simons2004,Codling2007,Torney2013}.   
For each of these benefits, individuals combine social and environmental information which enhances information processing\cite{Couzin2009}, possibly leading to emergent collective intelligence \cite{Woolley2010,Berdahl2013,Ioannou2016}.

However, social benefits are weighed against possible downsides\cite{Banerjee1992,Lux1995}, such as de-valued individual information\cite{Banerjee1992} and decreased sensitivity to changing environments\cite{Feldman1996}.  
Individuals must weigh environmental information gathered from their senses along with social information \cite{Couzin2011} or become isolated and face a greater risk of predation \cite{Krause2002,Handegard2012,Ward2016}.

Despite the large number of studies on collective behaviour, how individuals integrate social and environmental information across ecological contexts remains an open question.
Previous work using mixed societies of artificial and natural agents has shown group decisions arise from nonlinear feedback between local interactions between individuals \cite{Halloy2007}.
Using multi-robotic-fish systems\cite{Wang2017} and simulations\cite{Hein2015}, studies have explored the evolutionary mechanisms through which individuals use social information to better interpret noisy environmental information.

In a recent study, Berdahl et. al  examined how a school of golden shiners ({\it{Notemigonus crysoleucas})} collectively 
navigated a noisy environment ( a dynamic light field )\cite{Berdahl2013}.
While individual golden shiners could not detect the environmental gradient, the school was able to collectively swim toward darker waters.  
The emergent sensing arose from social interactions governed by a simple rule: golden shiners swim faster in bright regions and slower in dark regions.  
However, many animals, even those of microscopic scales such as bacteria~\cite{Thar2003}, are able to sense environmental gradients.  

In this work, we investigate the interplay of social and environmental information and how it impacts group performance.  
Building on recent work by Berdahl et al. \cite{Berdahl2013}, our study combines empirical data with numerical simulations to examine the performance of schools of fish in navigating dynamic light fields. We experimentally examined group gradient sensing ability in two different species of freshwater fish: golden shiners ({\it{Notemigonus crysoleucas})}  and rummy nose tetra ({\it{Hemigrammus bleheri}}).
While we find golden shiners are not able to sense the environmental gradient but are able to collectively find darker regions using an emergent sensing as in agreement with previous work\cite{Berdahl2013}, the tetras out-performed the shiners for all group sizes, which can be attributed to tetra's ability to sense the light gradient individually.
We propose a model based on the light intensity dependent speed-modulation proposed in the Berdahl-Couzin model \cite{Berdahl2013}.
Our model includes an additional gradient sensing term that can be tuned to investigate the interplay of environmental and social information and its effect on group performance.

Our results show that while an individual's ability to sense gradients generally improves the group's performance, there are downsides.  
With greater weight given to environmental information, individuals rely less on social information leading to larger nearest neighbour distance, eventually fragmenting the school. However, by balancing social and environmental information, the nearest neighbour distance can be minimised, decreasing predation risk \cite{Pitcher1993,Tien2004,Ward2016} while keeping group performance near optimal.
The relative gradient sensing weight that minimises nearest neighbour distance avoids relying too much on one source of information and produces simulations that agree with our experimental data for rummy nose tetras.

\begin{figure}[!tb]

\includegraphics[width=0.95\linewidth]{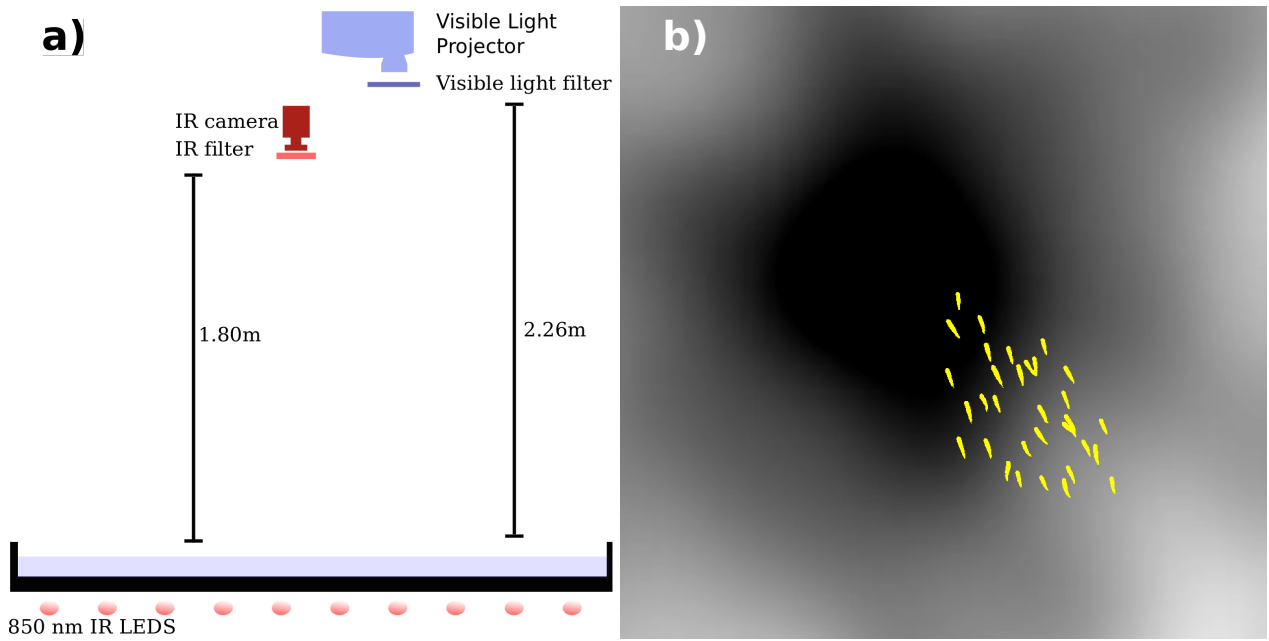}

\caption{ (Left) A schematic of the apparatus showing the infrared camera and projector placed overhead the experimental tank.  The tank is lit from below with several infrared lights.  (Right) The silhouettes of a group of 32 tetras are superimposed on the dynamic light field through
which they navigate.  The image is cropped, showing only a small region of the larger tank to illustrate the scale of dark spot to the body length of a fish.  }
\label{fig:Expt}
\end{figure}

\section*{Results}

\subsection*{Experimental results}
We filmed schooling events of two freshwater species, golden shiners ({\it{Notemigonus crysoleucas}}) and rummy nose tetras ({\it{Hemigrammus bleheri}}), in a shallow tank (183 cm $\times$ 102 cm, 8 cm water depth).   
As shown in \fref{fig:Expt}a, a projector located 226 cm over the experimental arena casts a dynamic light field at 30Hz onto the bottom of the tank.   
An IR camera placed $180$~cm over the tank records images at 30Hz as shown in \fref{fig:Expt}.
A cropped sample image of the light is shown in \fref{fig:Expt}b, with an overlay of the silhouettes of tetras for scale.  
Each noise image is the sum of a circular dark spot with gaussian decay ( with length scale 38.1 cm ) to white and a noisy greyscale light field that varies spatiotemporally \cite{Garcia-Ojalvo1992,Berdahl2013}.
The dark spot moves with a constant speed of 5.7 cm/s in random directions. 
The noise level $\eta = 0.25$ was held constant throughout the experiments.  
See the Supplementary Information and Berdahl et al. \cite{Berdahl2013} for further details on the light field.

We investigated the gradient tracking performance of schools of $N =$ 16, 32, 64, and 128 individuals.  
For each species and group size, we conducted five replicate experiments with different random seeds used to generate the light fields.  
Each experiment consisted of fish navigating the dynamic light field for 5 minutes.
Individual fish were tracked following a similar technique to Rosenthal et al. \cite{Rosenthal2015} to obtain trajectories of individual's positions.  
Velocities and accelerations were computed by convolving the position time-series with the first and second derivatives of a Gaussian, respectively \cite{Mordant2004}.  More details on our experimental methods and husbandry procedures are given in the Supplemental Information.

\begin{figure}[!tb]
\begin{center}
\includegraphics[width=0.95\linewidth]{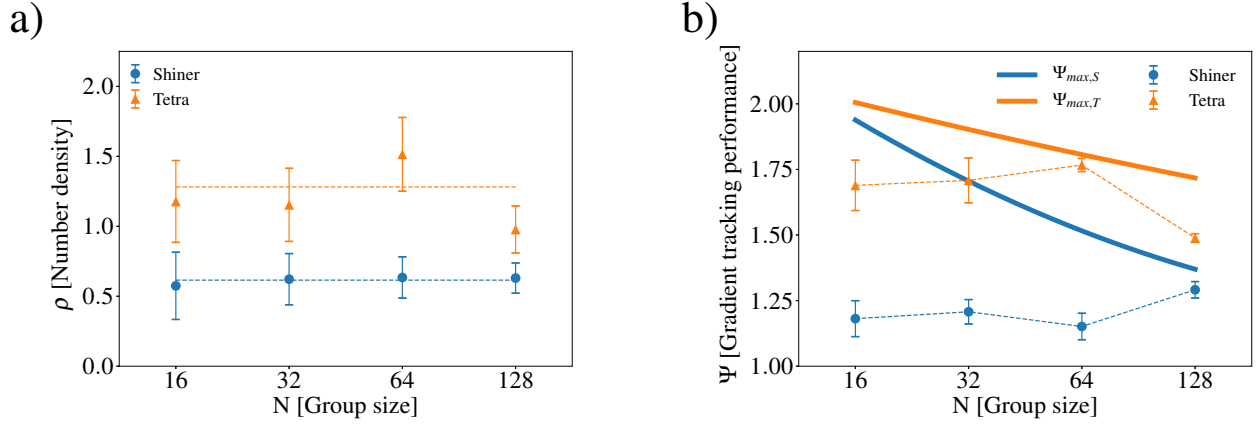}
\end{center}
\caption{(a) The average number density $\rho = A_\text{group} / (N  \cdot BL^2)$ shown for shiners (blue circles) and tetras (orange triangles) as a function of the group size N.  (b) The group gradient tracking performance, $\Psi$, is shown as a function of group size $N$ for both shiners (blue circles) and tetras (orange triangles).  The error bars represent the standard error of the group performance over replicates.  The thick lines represent maximum group performance of a group of $N$ individuals with the average area per individual for shiners and tetras. }
\label{fig:psiN}
\end{figure}

To quantify group gradient tracking performance, we calculate the mean gradient tracking performance as, $\psi = \langle  {\langle 1 - L \rangle_\text{fish}} \rangle_t$, 
which averages the local light level first over all fish in each frame and then over all time\cite{Berdahl2013}.  
This raw performance metric is then divided by the null performance, $\psi_{null}$, which is defined similarly
except by averaging over the level of darkness of fish trajectories if they instead experienced the temporal average of the light field.  
The un-biased gradient tracking performance is then,
\begin{align}
\Psi &= \psi / \psi_\text{null} .
\label{eqn:psi}
\end{align}

The theoretical maximum performance for a single individual is obtained by minimising $\Psi_{null}$ while remaining in the dark patch for the entire trial.  
Averaging over all five light fields generated with different random seeds, we calculate $\Psi_{max}  = 2.32  \pm 0.72$.   
While $\Psi_{max}$ gives a hard upper limit on group performance, this inaccurately assumes that fish can minimise $\Psi_{null}$ (fish have no knowledge of the temporal average of the light fields) and more importantly that fish occupy no volume.  
In \fref{fig:psiN}a, we show the number density $\rho = A_\text{group} / (N  \cdot BL^2)$  as a function of group size N for both shiners and tetras.
In all our experimental statistical analysis, data was analysed with R version 3.4.3 using a generalised linear mixed model (GLMM) approach with gamma errors using the lmer4 package\cite{Bates2014}, where light field random seed and home tank  were included as random factors to test for any effects with light field and group identity, respectively.
There was no significant effect of group size on the number density $\rho$ for both shiners  ($\chi^2 = 0.210$, $p=0.647$) and tetras ($\chi^2 = 1.768$, $p=0.184$).
The number density was $\rho = 0.69$ for shiners and $1.28$ for tetras, so the tetras are forming schools with roughly twice the number density as shiners.


In all our experiments, the radius of the circular dark spot is constant.  
Therefore, even if a school can perfectly track the spot, we expect that the group performance $\Psi$ decreases for schools that are sufficiently large as some individuals will be beyond the radius of the dark spot.
Note, however, that there are three factors which contribute to the area of the school: the number of fish $N$, the number density $\rho$ and the size of fish $BL$.
Firstly, the number density $\rho$ for both shiners and tetras is intensive (does not depend on N) as shown in \fref{fig:psiN}a, so increasing the number of fish $N$ linearly increases the overall area occupied by the school.  
Secondly, schools with lower number density $\rho$ occupy more area as the distance between each fish is larger.  
In \fref{fig:psiN}a, we showed that a school of tetras has roughly twice the number of density $\rho$ compared to shiners, therefore, tetras are forming denser, smaller schools than shiners.
Finally, for a given $N$ and $\rho$ increasing the size of the fish increases the overall area of the school.
The average body length of our shiners ($5.3\pm0.5$ cm) is about 50\% longer than the tetras ($3.4 \pm 0.5$ cm).
Shiners are larger fish that are forming less dense schools, therefore, we expect the optimal group performance for shiners to be less than tetras for large N.


For both shiners and tetras, the level of performance, $\Psi$ is shown as a function of group size in \fref{fig:psiN}b. 
To quantify how school size effects optimal performance for shiners and tetras, we calculate $\Psi_{max,S}$ and $\Psi_{max,T}$ for a circle of area $N$ times the average area occupied by a shiner or tetra, respectively.  
In \fref{fig:psiN}b, we show how $\Psi_{max,S}$ and $\Psi_{max,T}$ depend on N given the average area of a school for each respective species.
For small schools, $\Psi_{max}$ is large as the area of the school is smaller than the area of the dark spot.
However, as N grows, $\Psi_{max}$ decays with N, albeit decaying faster for shiners than tetras due to their larger area schools for a given N.

 The group performance of shiners was found to be independent of group size ($\chi^2 = 1.833$, $p=0.176$), which did not agree with the result found by Berdahl et. al\cite{Berdahl2013}. 
 Since larger schools occupy greater area, larger group size decreases the optimal group performance of a given size school $\Psi_{max,S}$.  
 In \fref{fig:psiN}b, note that the $\Psi(N=128) \approx \Psi_{max,S}(N=128)$ showing that large schools of  shiners are performing near optimally.
The slightly larger size of our shiners compared to those used in Berdahl et. al (BL = 4.9 cm)\cite{Berdahl2013} is likely responsible for the small decrease in performance for large schools of shiners.
Therefore, with smaller shiners (or a larger central dark spot), we expect our results to agree with Berdahl et. al.

Additionally, we found that group performance for tetras decreased with increasing group size ($\chi^2 = 4.620$, $p=0.032$).
However, we note that the group performance for N=128 is bounded by the decreasing optimal performance $\Psi_{max,T}$ which is due to the increasing area of the school but constant radius of the dark spot. 
While larger school area limits the group performance at large N, we found that tetras outperform shiners for small group sizes where the area of the school is small compared to the area of the central dark spot.  
We examine correlations between individual's acceleration to uncover the mechanism for the greater performance of tetras over shiners.

\begin{figure}[!tb]
\includegraphics[width=0.995\linewidth]{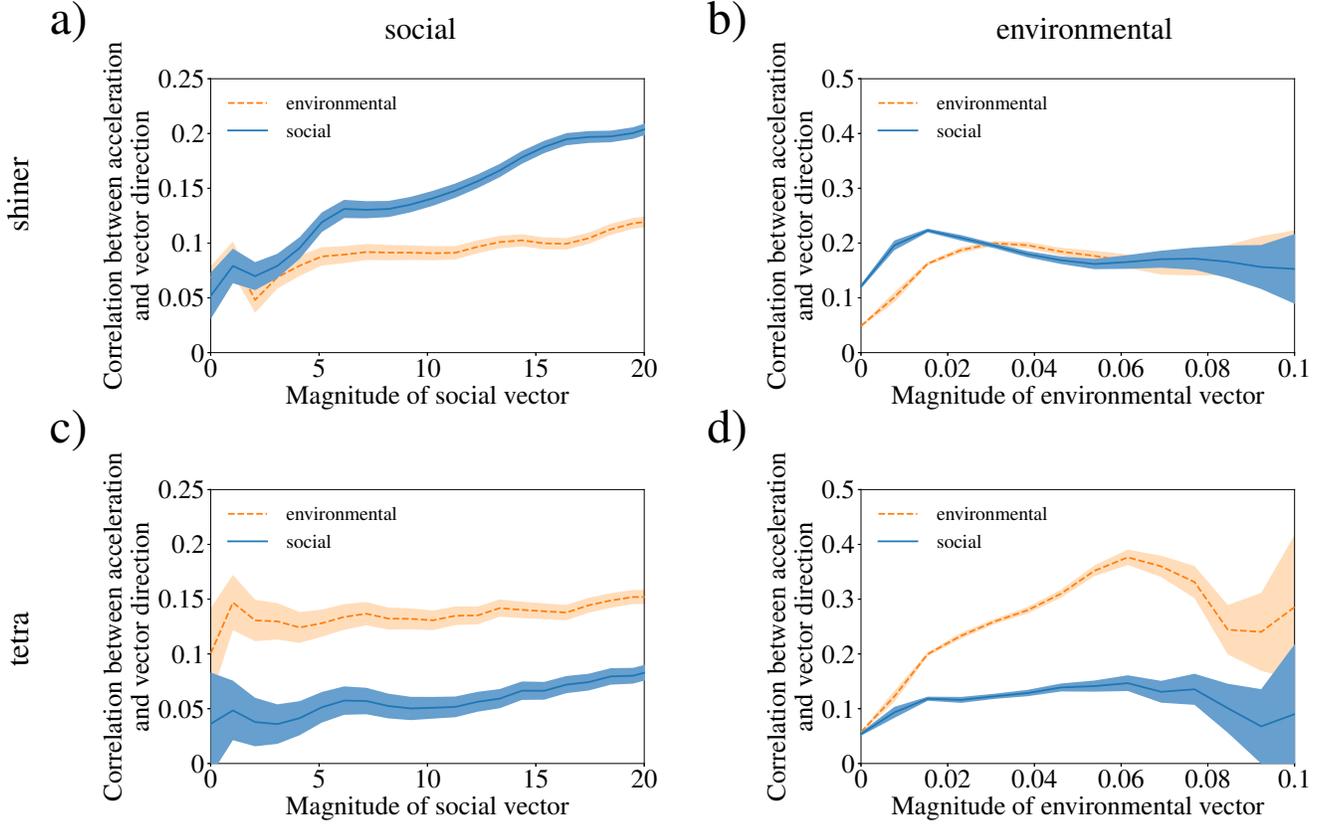} 
\caption{ 
Correlations between accelerations of shiners and the social and environmental cues (y-axis) are shown as functions of the magnitude of the social (a) and environmental (b) vectors (x-axis), respectively. Similarly, correlations between social and environmental vectors and the accelerations of tetras are shown as functions of the magnitude of the social (c) and environmental vectors (d), respectively.  
For all subfigures, the correlation between the accelerations and social vector are dark (blue) and between the accelerations and environmental vector are light (orange).  All shaded regions denote twice the standard error.
}
\label{fig:SocialVector}
\end{figure}

We investigated the mechanism for the increase in gradient tracking performance of tetras over shiners by examining the correlation between individual's acceleration and the social and environmental vectors.
We estimate a social vector $\vec{S}$ which is calculated using neighbours within seven body lengths of the focal individual,

\begin{align}
\vec{S}_i = \sum_{j\in r_s, j \neq i} \dfrac{ \vec{x}_j -   \vec{x}_i }{\vert \vec{x}_j -   \vec{x}_i \vert }.
\label{eqn:social}
\end{align}
\noindent The direction of the social vector indicates the direction of social attraction and its length is a proxy for the strength of the attraction.
While \fref{fig:SocialVector} uses seven body lengths to find $\vec{S}$, our results do not qualitatively change if the interaction range $r_s$ is between five and nine body lengths, as shown in Supplemental Figures S4 and S5.

We take the environmental vector $\vec{G}_i$ to be the negative gradient of the light field $L$ evaluated at the position $\vec{x}_i$ of each fish,

\begin{align}
\vec{G}_i = - \nabla L \vert_{\vec{x}_i}
\label{eqn:gradient}
\end{align}
\noindent The environmental vector $\vec{G}_i$ points in the direction of steepest descent toward darkness and its length is the rate of change of the light field in that direction.
To calculate the response of an individual fish to its social and environmental vectors, we calculate the correlation between the direction of 
the corresponding vector and the direction of the fish's acceleration with the following,

\begin{align}
 C_\text{social} &= \langle  \hatt{S}_i \cdot \hatt{a}_i \rangle \\
 C_\text{environmental} &= \langle  \hatt{G}_i \cdot \hatt{a}_i \rangle .
\label{eqn:corr}
\end{align}

Using $C_\text{social}$ and $C_\text{environmental}$, we determine whether the motion of individuals is more strongly correlated with social or environmental information.  

In \fref{fig:SocialVector}a, we show $C_\text{social}$ and $C_\text{environmental}$ for shiners as functions of the magnitude of the social vector $\vec{S}$.  
Since, $\vec{G}$ is independent of $\vec{S}$ we use the $C_\text{environmental}$ as a baseline to check the significance of $C_\text{social}$.  
For shiners, we find that $C_\text{social} \approx C_\text{environmental}$ when the magnitude of the social vector is less than 3, signifying that an individual's acceleration is poorly correlated with the social vector.
Note that the social vector's magnitude is determined by the number of individuals in the range and their spatial distribution.  
The magnitude of the social vector $\vert \vec{S} \vert$ is small when either few neighbouring fish are located within the interaction range or neighboring fish are located uniformly around the focal fish.
We do not expect shiners to respond strongly to the social vector in these situations.  
The magnitude of the social vector $\vert \vec{S} \vert$ is large when several neighbouring fish are located in a consistent direction.
In \fref{fig:SocialVector}a,  when $\vert \vec{S} \vert>3$, we find that $C_\text{social} > C_\text{environmental}$ and $C_\text{social}$ grows linearly with the magnitude of the social vector, signifying that the shiner's accelerations are correlated with the social vector.

In contrast, as shown in \fref{fig:SocialVector}b, $C_\text{environmental} \leq C_\text{social}$ for all magnitudes of the environmental vector, suggesting that individual shiners may not be able sense the environmental gradient, 
in agreement with recently reported results \cite{Berdahl2013}. 
When the magnitude of the environmental vector is very large, the uncertainty in $C_\text{environmental}$ and $C_\text{social}$ grows rapidly as the data is sparse and individual behavior is more variable.

In contrast to the data reported for shiners, we find that the accelerations of tetras are not strongly correlated with the social vector as shown in \fref{fig:SocialVector}c, as $C_\text{social} < C_\text{environmental}$.  
However, as shown in \fref{fig:SocialVector}d, tetras respond strongly to the light gradient as $C_\text{environmental}>C_\text{social}$ and $C_\text{environmental}$ increases with the magnitude of the environmental vector.
Note that this does not imply that tetras do not respond to social forces, only that their accelerations are more strongly correlated with the environmental vector.

While our results for shiners show that individuals are more strongly influenced by social information, tetras appear to exhibit the opposite trend and are more strongly influenced by the environmental gradient, as shown by $C_\text{environmental} > C_\text{social} $ shown in \fref{fig:SocialVector}d.
This result clarifies the findings in \fref{fig:psiN}b, where tetras outperformed shiners in navigating the dynamic light field toward darker regions.  
This stark contrast in performance between tetras and shiners in solving the same problem is due to the different gradient sensing mechanisms.
Shiners rely strongly on social cues which leads to an emergent group level gradient sensing\cite{Berdahl2013}, but tetras can individually sense the environmental gradient.

\subsection*{Simulation model}
As we have seen, different species of fish vary in the degree to which they base their movement on social and environmental information.  
To investigate this relationship, we explore how group performance depends on the relative weighing of environmental and social information via a tunable weight parameter, $w$.

\begin{figure*}[!ht]
\begin{center}
\includegraphics[width=0.995\linewidth]{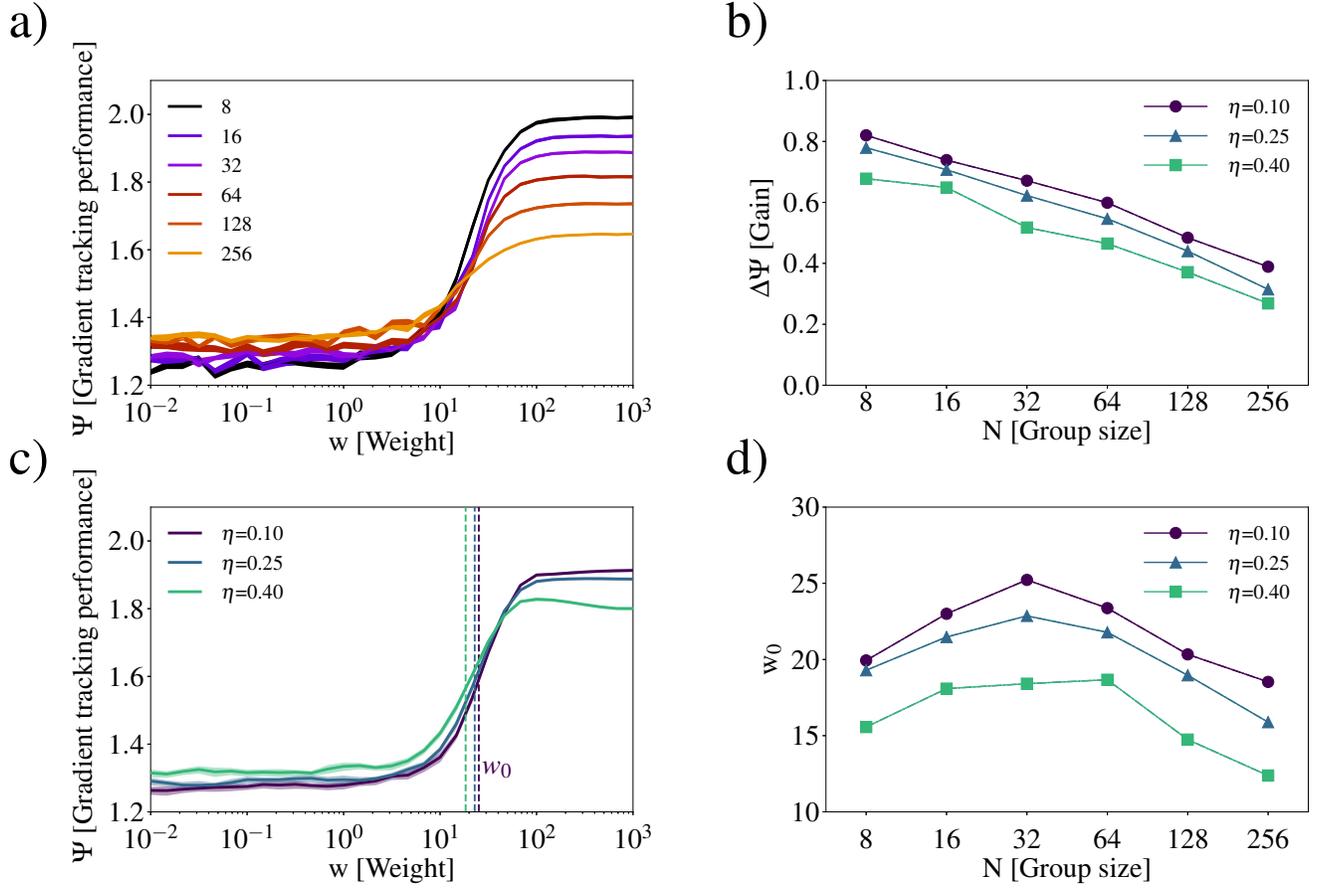} 
\end{center}
\caption{ (a) 
Group performance shown as a function of weight $w$ for group size $N$ = 8, 16, 32, 64, 128, and 256.  
(b) The performance gain $\Delta \Psi = \Psi_{max}(w) - \Psi(w=0)$ shown as a function of group size for different noise scales $\eta = 0.10, 0.25$ and 0.40. 
(c) Gradient tracking performance $\Psi$ shown as a function of weight $w$ for different $\eta$ and $N=32$.   
The dashed vertical line shows $w_0$ the weight at which $\Psi$ reaches half maximum, where $\Psi = \Psi (w=0) + 0.5 \Delta \Psi$.  
(d) The $w_0$ shown as a function of $N$ for different noise scales $\eta$.
}
\label{fig:weightData}
\end{figure*}

We propose a new model, built on the recently proposed agent based Berdahl-Couzin model\cite{Couzin2002,Couzin2005,Berdahl2013}, to which we augment with an environmental gradient sensing for each individual.  
Individuals interact socially via the canonical Couzin model where motion is determined via repulsive, aligning, and attractive interactions that depend on distance to neighbours.  
The direction given by social cues is $\hatt{d}_\text{social}$.  Explicit details of the calculation of $\hatt{d}_\text{social}$ are outlined in the Supplemental Information. We calculate the gradient of the noisy light field $\vec{d}_\text{environmental} = - \nabla L \vert_{\vec{x}_i} $.

The updated direction of each individual is given by 
\begin{align}
 \vec{d} = \hatt{d}_\text{social} + w ~\hatt{d}_\text{environmental},
\label{eqn:weight}
\end{align}
\noindent where $w$ is the relative weight and $ \hatt{d}_\text{social}$ and $\hatt{d}_\text{environmental}$ are unit vectors corresponding to the social and environmental vectors, respectively.
To determine the velocity of each individual, we then normalise $\vec{d}$ and multiply by the speed $s$,
\begin{align}
 \vec{v} = s~\hatt{d}
\label{eqn:velocity}
\end{align}

To determine the speed of individuals, we follow the Berdahl-Couzin model using a light intensity dependent speed-modulation proposed by Berdahl et al \cite{Berdahl2013}, where individuals slow down in dark regions and speed up in bright regions.  
The speed grows linearly with the brightness, given by  $s = s_\text{min} + L  \left( s_\text{max} - s_\text{min} \right)$.  Note, the speed modulation is the basis for the shiner's emergent sensing.  Further details and a flowchart for our model algorithm are given in the Supplementary Information.

In the limit that $w \rightarrow 0$, social interactions entirely determine individual behaviour and our model reverts to the Berdahl-Couzin model\cite{Berdahl2013}. 
In the opposing limit where $w \rightarrow \infty$, individuals lose all social information and respond only to their local environmental gradient.
We investigate the group performance as a function of group size $N = [8, ~16, ~32, ~64, ~128, ~256]$, the weight given to gradient information $w$ is 32 log-spaced values between $10^{-2}$ and $~10^{3}$, and the noise level of the environment $\eta= [0.10, ~0.25, ~0.40]$.  We performed 20 replicate simulations for each weight and noise level.  Each simulation was run for $10^4$ time steps and data was recorded every 100th time step. In all our simulations presented in the main text, the parameters used for the Couzin model \cite{Couzin2002} remained fixed throughout the simulations and
were: zone of repulsion 0.5; zone of orientation 3.0; zone of attraction 5.5; field of perception 270 degrees; turning rate 100 degrees; social error 0.01 radians; time step increment 0.125. 
These values correspond to the parameters fit for golden shiners as previously reported by Berdahl et. al \cite{Berdahl2013}.

\subsection*{Simulation results}
In \fref{fig:weightData}a, we show the group performance $\Psi$ of our simulation as a function of the weight $w$ with a  noise level $\eta = 0.25$.  The shape of $\Psi(w)$ is  sigmoidal.
We find that increasing $w$ (the gradient sensing weight)  greatly increases $\Psi$ of the group.  
In this limit, individuals follow the gradient eventually finding the moving dark region.
However, inside the dark region $\vec{d}_\text{environmental} = 0$, since the gradient is zero, and the motion of individuals is determined by social interactions.
Increasing group size $N$ decreases the maximum $\Psi$, which is due to the increasing size of the school.
Therefore, larger group sizes do not benefit from large gradient sensing weight as much as smaller groups.  
To quantify this effect, we calculate the max gain in group performance $\Delta \Psi = \max (\Psi) - \Psi(w=0)$.   
In \fref{fig:weightData}b, we show $\Delta \Psi$ as a function of group size, N for three different noise levels $\eta = 0.10, 0.25,$ and $0.40$.
We find that $\Delta \Psi$ decreases monotonically with increasing $N$ and decreases with increasing $\eta$, showing that larger groups benefit less from an individual's ability to sense the gradient regardless of the noise level of the environment.

\begin{figure}[!b]
\begin{center}
\includegraphics[width=0.65\linewidth]{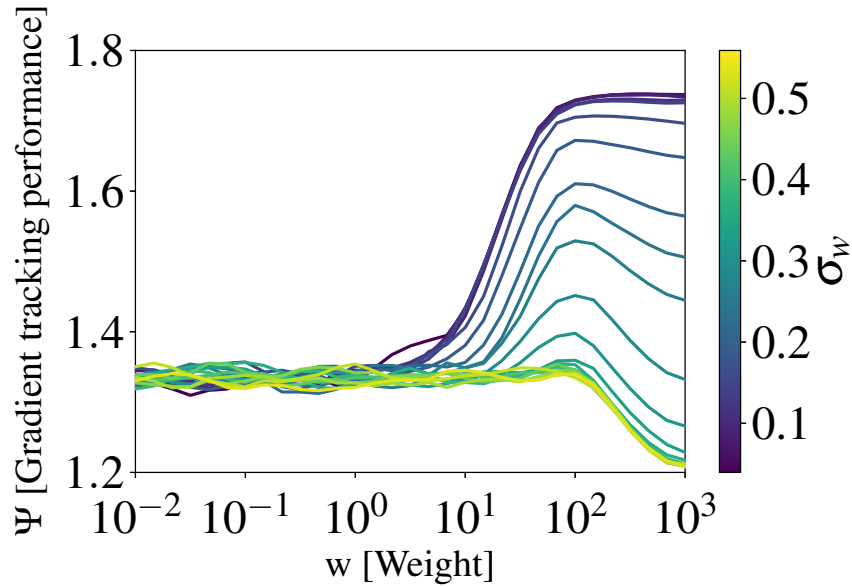}
\end{center}
\caption{The gradient tracking performance $\Psi$ is shown for $N=128$ as a function of weight $w$ for different gradient sensing error $\sigma_w$.}
\label{fig:weightError}
\end{figure}

The group performance decreases  with increasing noise level $\eta$ at large $w$, as shown in \fref{fig:weightData}c.  In more noisy environments, individuals who strongly rely on gradient sensing benefit less, as they respond to local variations in light level instead of large scale features.

Another trend of increasing $\eta$ is the shifting of the sigmoidal $\Psi(w)$ toward smaller $w$.
The sigmoid reaches half the max, $\Psi_{1/2} = \Psi(w=0) + \frac{1}{2} \Delta \Psi$, at $w_0$ as shown in \fref{fig:weightData}c.  
In \fref{fig:weightData}d, we show the shifting of the group performance $\Psi$ toward smaller $w$ by showing $w_0$ as a function of $N$ and $\eta$.  
The weight at half-max, $w_0$, decreases with increasing environmental noise level  $\eta$.  
We also find a maximum of $w_0$ for groups of $N=32$, demonstrating that small (and large) group benefit more from smaller weights $w$ than intermediate group sizes.

Up to this point, our study did not account for the ubiquitous presence of error in individual environmental gradient sensing. 
To introduce this effect, we include a gaussian random error with standard deviation $\sigma_w$ which is added at each timestep to the local gradient direction.  
As shown in \fref{fig:weightError}, for small $w$, the amount of error added to the gradient direction is nullified as individuals base their direction on social interactions.  
For large weights, $w>100$, increasing the amount of error $\sigma_w$ destroys the individual ability to sense the gradient and turns the simulation into non-interacting random walkers.

\begin{figure*}[!tb] 
\begin{center}
\includegraphics[width=0.995\linewidth]{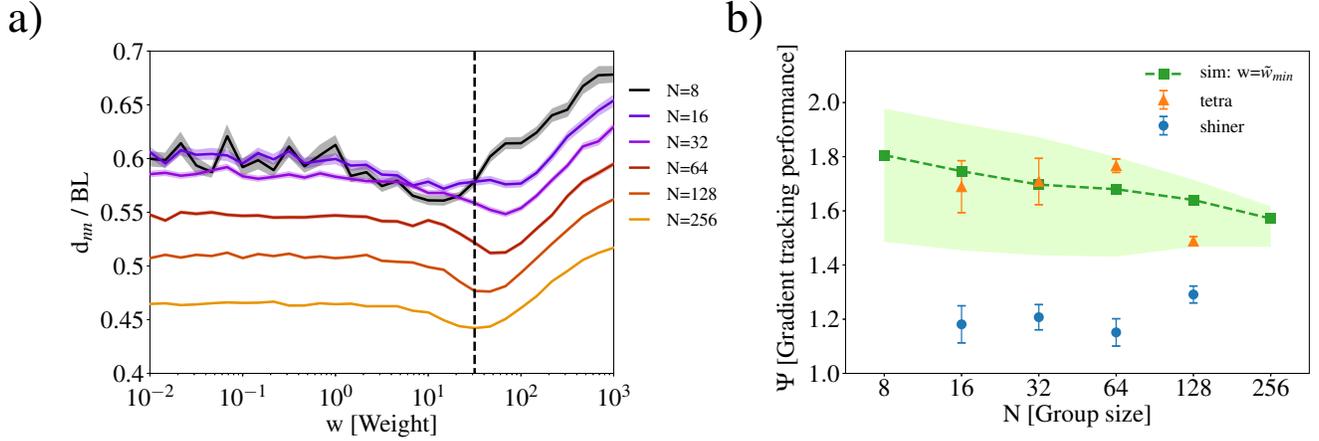} 
\end{center}
\caption{ (a) 
Nearest neighbour distance of simulated schools is shown as a function of the weight $w$ for group sizes $N$. The dashed vertical line is the weight $\tilde{w}_\text{min}$ which minimises $d_{nn}$ for all $N$.
(b) Gradient tracking performance of the numerical results for $w = \tilde{w}_\text{min} \approx 32$ (dark green squares), where the shaded region shows the range of $\Psi$ corresponding to simulated weights from $w \approx 14$ to $64$. These weights are those which minimise $d_{nn}$ for $N = 256$ and $16$ respectively.
The experimental results are overlaid for the rummy nose tetras (orange triangles) and shiners (blue circles).}
\label{fig:weightNND}
\end{figure*}

While increasing $w$ increases the group performance $\Psi$, large $w$ destroys social information.  
To quantify this effect, we calculate the nearest neighbour distance $d_{nn}$. 
For all group sizes $N$, nearest neighbour distance decreases with increasing $w$ to a minimum $\text{min} (d^N_{nn} )$.  
We find that $d_{nn}$ decreases for increasing group size $N$, corresponding to the increasing density of the group near the dark spot.
Note that $d_{nn} < 0.5 \text{BL}$ for $N\geq 128$ which is due to overcrowding in the dark spot and low time resolution.  
With shorter $\Delta t$ for the simulation, the repulsive social interaction would prohibit such overcrowding and should yield a minimum $d_{nn} \geq 0.5 \text{BL}$. 
For large $w$, we find that $d_{nn}$ increases, as long range attractive social interactions weaken and individuals follow local transient gradients away from the school.

We propose that nature selects the weight $w$ which balances individually acquired gradient information with social information such that nearest neighbour distance $d_{nn}$ is minimised for all group sizes.  
Smaller nearest neighbour distance reduces predation risk \cite{Pitcher1993,Tien2004,Ward2016}. 
We find the weight which minimises the sum of the squared difference of $d_{nn}$ for group sizes $N = 16, ~32, ~64$ and 128 to be $\tilde{w}_\text{min} = 31.6$, and is shown in \fref{fig:weightNND}a.
Therefore, by weighing their individually acquired gradient information around thirty times stronger than social interaction, individuals can optimise both $\Psi$ and $d_{nn}$.
However, we emphasise the granularity of $w$ due to computational cost.
Using $w=\widetilde{w}_\text{min}$, we show the group performance $\Psi$ as a function of $N$ in \fref{fig:weightNND}b.
We find good agreement between our model and experimental data for the rummy nose tetras.  
In \fref{fig:weightNND}b, the shaded region represents the $\Psi$ for the range of $w$ that minimise $d_{nn}$ in simulations for each group size $N$, $w = 14.7$ to $68.1$.
Therefore, to further improve the model's ability to match the tetra's behaviour, one could optimise $r_o$, $r_a$ and $w$ to better replicate the behaviour of the tetra, however, given the large error bars on the simulated group performance and the good agreement with experimental data, parameterising the data for the tetras may be of limited value.  
We find that the weight which minimises $d_{nn}$ also minimises the root mean square error (RMSE) in group performance for experimental data of the tetras.
For groups $N=128$, the group performance for $\Psi$ for the tetras falls below the range predicted by our simulation.  
We attribute this disagreement largely to distraction of individual fish with the edge of the tank.


Note that the simulations are not parametrised to fit the schools of tetra.   
In all the numerical results, we fixed the repulsion, orientation and attraction zonal distances to match those reported for golden shiners previously reported \cite{Berdahl2013}.  
We use the simulations to display a generic behaviour of self-propelled particle models which supports our experimental observations that tetras have some innate environmental gradient sensing ability.
In the Supplemental Information Fig. S8, we show that the minimum of the RMSE between the tetras and the performance for our model using $w=0$ (Berdahl-Couzin) gives a smaller interaction range ($r_o, r_a$), but the RMSE is several times larger than that reported for the shiners \cite{Berdahl2013}. 
Furthermore, using this $r_o$ and $r_a$, we conduct a parallel analysis as in the main text and find that the agreement with the experimental data was not as good as results reported in the main text in \fref{fig:weightNND}b.
We refer the reader to the Supplemental Information for further discussion.


\section*{Discussion}

This work presents experimental and numerical studies on group performance of fish schools navigating a spatiotemporally varying light field.  
First, we experimentally investigated the collective gradient tracking performance of two freshwater species: golden shiners ({\it N. crysoleucas}) and rummy nose tetras ({\it H. bleheri}).
Our results agree with previous findings that the motion of shiners is based strongly on social interactions and is not correlated with the light gradient \cite{Berdahl2013}.
However, our results show that tetras outperform shiners at this task for all group sizes, which is due to the individual ability for tetras to sense the gradient of the light field. 
The emergent collective sensing (shiners) and collective enhanced individual sensing (tetras) are distinct gradient sensing mechanisms.

Second, we used these observations to propose an agent-based model based on the Berdahl-Couzin model~\cite{Couzin2002,Berdahl2013}, where we included a gradient sensing ability which is weighted against social information for each individual.
In our simulations, we find that group performance increases with increasing dependence (weight) on gradient sensing information, which is robust to group size $N$ and noise scale $\eta$ as shown in \fref{fig:weightData}ac.
However, when individuals rely too much on their individual information, social information is lost, nearest neighbour distance increases and the school fragments.

To balance the benefits of social information and group living\cite{Krause2002,Ward2016} with individually acquired environmental information, we
proposed that individuals adjust the relative weight attributed to environmental (gradient) or social information to minimise the average nearest neighbour distance $d_{nn}$.
This adaptation is important, as decreasing nearest neighbour distance reduces risk of predation \cite{Pitcher1993,Tien2004,Ward2016}.
Using the gradient sensing weight which minimises the nearest neighbour distance, we found good agreement with our experimental data for tetras.   
Furthermore, we found that the minima in the $d_{nn}$ depends on $N$, which suggests that it may be advantageous for individuals to dynamically adjust the weight based on group size.  
A dynamic tuning based on the situation would allow individuals to optimise their performance based on the situation or quality of information.
For example, a dynamic tuning would allow individuals to favour individually acquired environmental information, when the environment is less noisy or the magnitude of the environmental vector is large.  
By contrast, if the magnitude of the social vector is large or the environment is noisy\cite{Lihoreau2017}, individuals can benefit from decreasing the weight and relying more on social information.

Dynamic and adaptive behavioural rules have been experimentally investigated in groups of fish exposed to alarm or food cues.  
Under predatory cues, x-ray tetras slow down and increase in density\cite{Schaerf2017} and bluegill sunfish form more highly polarised schools\cite{Ioannou2012}.   
Fish swim faster and decrease in density when exposed to food\cite{Hoare2004,Schaerf2017,Miller2007}.
While individual differences in physical ability and sensing can have large implications on decision making\cite{Ward2011} and the fitness of the group\cite{Jolles2017}, social benefits may pressure individuals to ignore private information and conform\cite{Herbert-Read2013}.
In these situations, individuals adjust their social interactions via rapid feedback loops in order to conform to the group\cite{Harcourt2009,Farine2015,Jolles2017}.


For rummy nose tetras, we found group performance decreases with group size, which is due to geometrical constraints as size of the school exceeds the size of the dark spot.
Our results show that tetras can maximise their benefit in smaller groups.
In some contexts, larger groups may improve the group performance\cite{Berdahl2013} and facilitate fast and accurate decisions where there are predatory cues\cite{Ward2011}, though larger groups can increase confidence without increasing decision accuracy\cite{Lorenz2011,Kao2014} and in some cases deteriorate obstacle avoidance capabilities\cite{Gelblum2015}.
As often seen in nature\cite{Krause2002,Sumpter2010}, individuals may maximise their benefit in small to intermediate sized groups with increased decision accuracy from inherent noise and less detrimental feedback from correlated information (present in larger groups)\cite{Kao2014}.
In {\it P. longicornis} ants where the food is collectively transported, the collective response is maximised for intermediate group sizes where criticality facilitates the flow of new information\cite{Gelblum2015}.
In a similar experiment to the one in the text and Berdahl et. al\cite{Berdahl2013}, however using humans, collective sensing was maximised in smaller groups, where collective sensing strategies were learned in minutes\cite{Krafft2015}.



In summary, our model shows the disadvantages of extreme behavioural rules, where either relying too strongly on either social information (w=0) produces in poor group performance or relying too strongly on individually acquired environmental gradient information ($w\rightarrow \infty$) destroys valuable social information and fragments groups.
Our results show that individuals can balance their individually acquired environmental information with social information, which promotes group performance and strong cohesion.
Our experimental results show two freshwater species using distinct gradient sensing mechanisms: emergent collective sensing (shiners) and collective enhanced individual sensing (tetras).
These interaction rules evolved under distinct ecological and social conditions\cite{Hein2015}.
In new environments, individuals value social information more than their individually acquired environmental information ($w<1$) which is due the need to conform\cite{Ward2011}.
Therefore, it is likely that individuals also adjust their weighing of social and environmental information $w$.
An extension of this work would explore the extent to which individuals tune $w$ to conform with the group or if $w$ depends on the environmental noise level.


\section*{Methods}

\subsection*{Husbandry}
We studied the gradient sensing performance of schools of golden shiners ({\it Notemigonus crysoleucas} ) and rummy nose tetras ({\it Hemigrammus bleheri}) in a laboratory.  While both are freshwater fish that prefer to school in dark shallow water, shiners (a cyprinid found in cool waters of eastern North America) and tetras (a characin found in the tropical waters of Amazon Basin of Brazil and Peru) require different water chemistry.

The golden shiners {\it N. crysoleucas} were acquired from Anderson Minnows.  
We kept approximately 500 juvenile shiners in three 30 gallon home tanks ($\approx150$ in each tank) using de-chlorinated, aerated, and filtered tap-water kept at $21^o$C.  
Water changes of 30\% were done twice weekly.  
Shiners were $5.3\pm0.5$cm in length.  

The rummy nose tetras, {\it H. bleheri}, were acquired from Cichlid Exchange.  We kept approximately 200 tetras in two 40 gallon home tanks (100 in each) at a constant temperature of $27.0 \pm 0.5^o$C in a 1:3 de-chlorinated tap water to reverse osmosis water that was aerated and filtered.  The RO water diluted the pH and gH of the tap water to $6.8 \pm 0.2$ and $100 \pm 20$ppm, respectively. 
Water changes of 20\% were done once a week.  
Tetras were $3.4\pm 0.5$ cm in length. 

The home tanks for both species were illuminated with 12 h of light and 12 h of darkness per day. 
Both were fed a mix of crushed TetraMin flakes, Hikari brine shrimp, micro pellets, and freeze-dried blood worms four times a day.  
Before each experimental trial, fish were gently netted from their home tanks an hour after their first feeding and transferred to the experimental tank.  
We ensured that fish were not used in experiments on consecutive days by using a rotating schedule to select which home tank to gather fish.
Fish were appropriately acclimatised to the water in the experimental tank before experiments took place.  
Further husbandry details are outlined in the Supplementary information.

\noindent\textbf{Ethics.} All experiments were conducted in accordance with federal and state regulations and were approved by the Gettysburg College Institutional Animal Care and Use Committee.

\subsection*{Experimental setup}

We conducted experiments with golden shiners and rummy nose tetra in a quasi two-dimensional acrylic tank ($183 \times 102$ cm, $8$ cm water depth). 
Videos of schooling events were captured via a USB3 Point Grey camera mounted $180$ cm above the tank which was back-lit by 850nm infrared LEDs positioned beneath the tank. 
The videos were captured in $2048 \times 1280$px at 30 frames per second by the camera which was hardware triggered to synchronise with the projected light field. 
The dynamic light field of $940 \times 540$px were generated by a projector positioned 226 cm above the experimental tank at 30 frames per second.  
As shown in Supplemental Figure S2 and zoomed in \fref{fig:Expt}, the projected field consisted of a single dark spot which moved randomly around the tank at a constant speed and was overlaid on a noisy background which varied both spatially and temporally, identical to the method detailed in a previous work\cite{Berdahl2013}.  
Measured light levels at the surface of the tank ranged from 10 lux (approximately twilight) to 500 lux (sunrise on a clear day), corresponding to the natural environment of the fish in the morning or evening. 

For each species (tetra and shiner) and group size $N = 16$, $32$, $64$ and $128$, we recorded five replicate experiments for 5 minutes.   
We used four different seeds to generate each projected video at a medium level of environmental noise ($\eta = 0.25$), and added a 50 pixel white border to the light field to discourage fish from interacting with the sides of the arena.  
Each experimental run was followed by a 10 minute rest period under neutral lighting ( 0.5 lux, deep twilight ).  
See Supplementary information for further experimental details.


\subsection*{Fish Tracking}

Our algorithm was implemented in Python using the OpenCV library, and followed a similar approach to SchoolTracker \cite{Rosenthal2015}.
Individual fish are located using detected line-segments in background subtracted frames of video.
We then track fish from frame to frame by linking their two-dimensional positions over time using a Kalman filter.  
Due to the large number of fish, occlusions are frequent and the detection/tracking algorithm can fail to locate and track a fish over multiple frames.
The tracks are spliced together by linking tracks in a four-dimensional position-velocity space \cite{Xu2008}.  
Our tracking algorithm recovers 90\% of trajectories for $N=128$ and 92\% for $N=16$.  
Since our focus in this paper does not rely on us maintaining identities for long periods of time, we have sufficient data to calculate velocities and accelerations.
  
Once the time-resolved trajectories are known, we compute velocities and accelerations by convolving the trajectories
with a Gaussian smoothing and differentiating kernel \cite{Mordant2004, Ouellette2007}.
Derivatives computed using this convolution method are less noisy than what would be obtained from a simple finite difference scheme. 
For the data presented here, the convolution kernel was chosen to have a standard deviation of 1.5 frames, and the position information from 11 frames was used to calculate each derivative.

\section*{Acknowledgements}
We want to thank the two anonymous reviewers whose suggestions greatly improved this manuscript.
We gratefully acknowledge the support of NVIDIA Corporation with the donation of the Titan Xp GPU used for this research. This work is also supported by Gettysburg College and by the Cross-Disciplinary Science Institute at Gettysburg College (X-SIG).

\section*{Author contributions statement}

J.P. devised the study.  A.P., J.G., and J.P authors conducted the experiments.  A.P. and J.P. wrote the simulation code and performed the simulations.  All authors analysed the data, interpreted the results and wrote the manuscript.

\section*{Additional information}
The authors declare no competing interests.



\end{document}